\documentclass[11pt,preprint,sort&compress,times]{elsarticle}
\usepackage[utf8]{inputenc}
\usepackage[usenames]{color}
\usepackage{textcomp}
\usepackage{amsmath}
\usepackage{graphicx}
\usepackage{float}
\PassOptionsToPackage{version=3}{mhchem}
\usepackage{mhchem}
\usepackage[unicode=true,
 bookmarks=false,
 breaklinks=false,pdfborder={0 0 1},backref=section,colorlinks=false]
 {hyperref}

\makeatletter

\usepackage{commath}
\usepackage{longtable}
\usepackage{tcolorbox}
\usepackage{siunitx}
\usepackage{physics}
\usepackage[displaymath]{lineno}

\definecolor{AL}{rgb}{1.0,0,0}
\definecolor{AM}{rgb}{0,0,1.0}

\newcommand{\iaa}{
IAA-CSIC, P.O. Box 3004, 18080 Granada, Spain}

\journal{Computer Physics Communications}

\makeatother

\begin{document}
\begin{frontmatter}
\title{A domain-decomposition method to implement electrostatic free boundary
conditions in the radial direction for electric discharges}

\author{A. Malag{ó}n-Romero}

\address{\iaa}

\ead{amaro@iaa.es}

\author{A. Luque}

\address{\iaa}
\begin{abstract}
At high pressure electric discharges typically grow as thin, elongated
filaments. In a numerical simulation this large aspect ratio should
ideally translate into a narrow, cylindrical computational domain that
envelops the discharge as closely as possible. However, the development
of the discharge is driven by electrostatic interactions and,
if the computational domain is not wide enough, the boundary
conditions imposed to the electrostatic potential on the external
boundary have a strong effect on the discharge. 
Most numerical codes
circumvent this problem by either using a wide computational domain
or by calculating the boundary conditions by integrating the Green's 
function of an infinite domain. Here we describe an accurate and efficient
method to impose free boundary conditions in the radial direction for an elongated electric discharge. To facilitate the use of our method we provide a sample implementation. Finally, we apply the method to solve Poisson's equation in cylindrical coordinates  with free boundary conditions in both radial and longitudinal directions. This case is of particular interest for the initial stages of discharges in long gaps or natural discharges in the atmosphere, where it is not practical to extend the simulation volume to be bounded by two electrodes. 
\end{abstract}


\end{frontmatter}

\newpage
\noindent
\textbf{PROGRAM SUMMARY} 

\begin{small} 
\noindent
 {\em Program Title: \verb|poisson_sparse_fft.py|} \\
 {\em Licensing provisions:} CC By 4.0 \\
 {\em Programming language:} Python \\

\noindent
{\em Nature of problem:}
Electric discharges are typically elongated and their optimal computational domain
has a large aspect ratio.  However, the electrostatic interactions within the
discharge volume may be affected by the boundary conditions imposed to the
Poisson equation.  Computing these boundary conditions using a direct 
integration of Green's function involves either heavy computations or a loss
of accuracy.\\
{\em Solution method:}
We use a Domain Decomposition Method to efficiently impose 
free boundary conditions to the Poisson equation.  This code provides
a stand-alone example implementation.\\
  \\

 ©2018. This manuscript version is made available under the CC-BY-NC-ND 4.0 license http://creativecommons.org/licenses/by-nc-nd/4.0/

 \end{small}

\section{Introduction}

Despite their prevalence in industry and in nature, electric discharges
still hold many unknowns. For example, we do not yet understand precisely
how a lightning channel starts, how it advances in its way to the
ground or how exactly are bursts of X-rays
produced as it progresses \cite{Dwyer2014/PhR}. This is partly due to 
the short time and
length scales involved in such processes which, combined with their
jittery behaviour, prevents the use of many diagnostic techniques.
Due to these limitations, much of what we know about electric discharges
comes from computer models which, at least within a simulation, are
predictable and reveal arbitrarily small scales.

Consider streamer simulations. Streamers are thin filaments of ionized
air that precede most electric discharges in long gaps at atmospheric
pressure. The main challenge for simulating streamers is the wide
separation between length scales: whereas the total length of the
streamer channel at atmospheric pressure ranges from about one to
some tens of centimeters, the ionization of air molecules is mostly
confined to a layer thinner than one millimeter. Despite this difficulty,
there are many numerical codes that explain most of the observed properties
of streamers \cite{Ebert2006/PSST,Liu2006/JPhD,Luque2012/JCoPh,Liu2015/NatCo,Qin2015/GeoRL,Teunissen2017/arXiv}.
In the past decades these models have gradually improved and successfully
overcome many of the challenges posed by streamer physics. However,
they are still computationally intensive and often require days of
runtime to produce meaningful simulations.

In this work we look at one of the problems behind these long
running times: the large aspect ratio of a single-channel discharge.
Whereas the width of an atmospheric-pressure streamer is at most about
one centimeter, its length spans many times this extension. In order
to minimize the amount of work performed in a simulation, one strives
to adapt the computational domain to the dimensions of the streamer,
which means using a narrow cylindrical domain with a diameter only
slightly larger than the streamer width. However, in such a narrow
domain the electrostatic interaction between separate points in the
channel is strongly affected by the boundary conditions imposed on
the electric potential at the outer boundaries.

One approach to avoid this artifact while keeping a narrow domain
around the streamer is to calculate the boundary values of the potential
by direct integration of the electrostatic Green's function in free
space \cite{Babaeva1996/JPhD,Babaeva2000/chapter,Liu2004/JGRA,Liu2006/JPhD,
Bourdon2007/PSST}. 
These values are then imposed as
inhomogeneous Dirichlet boundary conditions in the solution of the
Poisson equation. In a cartesian grid with $M$ cells in the radial
direction and $N$ cells in the axial direction the direct integration
of the Green's function at each of the $N$ nodes in the external
boundary requires about $MN^{2}$ operations. Since the work employed
by fast Poisson solvers scales as $MN\log(MN)$ ($MN$ for multigrid
solvers), the computation of boundary values by direct integration
may easily dominate the work employed in the electrostatic calculations.
This is mitigated in part by using a coarse-grained charge distribution
in the integration. However, in that case there is a tradeoff between
the degree of coarsening and the minimal radial extension of the domain
required for a tolerable error.

Beyond this common approach used to solve Poisson's equation in electric discharges, 
some other methods have been developed. 
A family of these methods has been built upon the idea of the
decoupling of local and far-field effects \cite{Anderson/JCP} 
and the computation of the boundary potential
by means of a potential generated by a set of screening charges
located in the outer surface of the computational domain \cite{James/JCP}.  
Based on these two methods mentioned above, 
reference \cite{Balls/JCP} uses a domain decomposition approach 
to exploit parallel computing capabilities; first, 
Poisson's equation subject to unbounded boundary conditions is solved in a set of disjoint patches. 
As a second step a coarse-grid representation
of the space charge is obtained and Poisson's equation is again solved in a global coarse-grid 
whose solution is used to communicate far-field effects
to local patches. Finally, Poisson's equation is solved in a fine grid 
using boundary conditions computed  from the coarse-grid solution 
corrected with local field information.

A different family of methods uses
the convolution with Green's function subject to free boundary conditions.
They manage the singular behaviour of Green's function
by either regularizing it  \cite{Hejlesen/JCP}, 
or by replacing the singular component to the integrand
of the convolution by an analytical contribution \cite{Anderson2016/JCP}. 
These methods have achieved an order of convergence greater than two.

Here we adapt to the cylindrical geometry of electric discharges the
domain-decomposition method described by Anderson 
\cite{Anderson1989/chapter} (see also \cite{Bayliss1982/SIAMJAM} for a review of
similar techniques). As we discuss below, this method requires
two calls to the Poisson solver but otherwise the leading term in
its algorithmic complexity follows the scaling of the Poisson solver
itself. Therefore for large grid sizes our approach is more efficient
than the direct integration method. Furthermore, as we do not reduce
the resolution, we do not introduce any numerical error in addition to
the discretization error of the Poisson equation. We believe that the method we present is simple enough that it can be easily
implemented on top of any existing streamer simulation code. To aid
in this task we provide a standalone example in Python.

Some applications may also require free boundary conditions for the 
$z$-direction: for 
example, when the discharge develops far from the electrodes.  In those cases one may also reduce the computational 
domain in the longitudinal direction while the core of the simulation remains inside the computational domain. We have considered 
this topic of interest in \ref{sect:fullfree} where we have applied the 
domain decomposition method to obtain free boundary 
conditions also in the longitudinal direction. This extension requires an extra solution of Poisson's equation.

Note that streamers are not the only type of discharge that typically
exhibits a large aspect ratio and that therefore our scheme is also
applicable to other processes such as leaders and arcs.
 
\section{Description of the method}

\label{sect:description} 

\subsection{Domain decomposition}

\begin{figure}
\includegraphics[width=0.5\columnwidth]{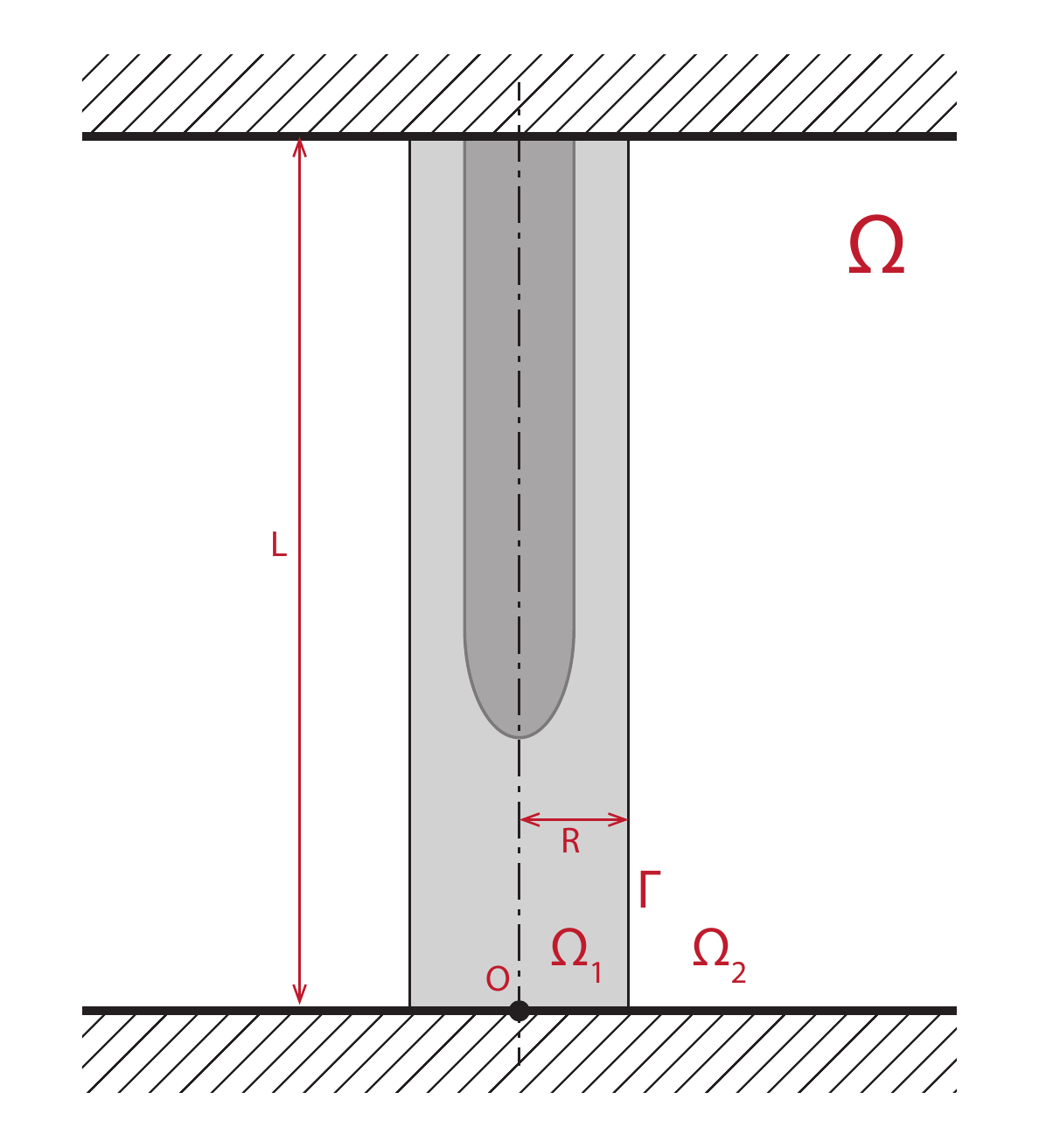} \caption{\label{fig:sketch} Geometry of the discharge considered in this work.
An elongated channel propagates between two conducting electrodes.
The space between these electrodes, $\Omega$ is divided into two
domains: the inner domain $\Omega_{1}$ is our computational domain
and contains all the space charge. The outer domain $\Omega_{2}$
extends indefinitely outwards from the external boundary of $\Omega_{1}$
and does not contain any space charge. The cylindrical surface $\Gamma$
is the common boundary between $\Omega_{1}$ and $\Omega_{2}$. }
\end{figure}

The most convenient decomposition of the domain strongly depends on the problem at hand.
The decomposition we present here is suitable for elongated discharges and probably
some other applications but the procedure and the highlighted ideas are not restricted to this particular
scheme.

We consider the geometry sketched in figure~\ref{fig:sketch}, where
an elongated, cylindrically symmetrical streamer propagates between
two planar electrodes. With minimal changes, our scheme can be extended
to more complex geometries commonly employed in streamer simulations,
such as protrusion-plane, protrusion-protrusion and sphere-plane.
The electrostatic potential $\phi$ satisfies the Poisson equation
with appropriate boundary conditions: 
\begin{equation}
\begin{array}{ccc}
\Delta\phi=f & \text{in} & \Omega,\\
\phi=g & \text{on} & \partial\Omega,
\end{array}\label{eq:Initial problem}
\end{equation}
where $f=-q/\epsilon_{0}$, with $q$ being the charge density and
$\epsilon_{0}$ the vacuum permittivity. In principle an arbitrary boundary 
condition, here denoted by $g$, can be applied to the upper and lower 
electrodes.  However, to simplify our discussion we limit ourselves to the most 
common case where
$g=0$, meaning $\phi=0$ at $z=0$ and $z=L$ (to impose
a potential difference $V$ between the two electrodes we simply add
$\phi_{\text{inhom}}=zV/L$ to the solution of the homogeneous problem).
The domain $\Omega$ is the space between the two electrodes, formally
defined as 
\begin{equation}
\Omega=\left\{ x\equiv\left(\rho,\theta,z\right)\in\mathbb{R}^{3}/0\leq\rho,0\leq\theta<2\pi,0\leq z\leq L\right\} .
\end{equation}
Since our geometry is cylindrically symmetrical, we will henceforth
omit the variable $\theta$ and consider the two-dimensional domain
spanned by the variables $(\rho,z)$.

Our purpose is to decompose the physical domain $\Omega$ into two,
which we name $\Omega_{1}$ and $\Omega_{2}$, such that $\Omega=\overline{\Omega}_{1}\cup\overline{\Omega}_{2}$
($\overline{\Omega}_{i}$ is the closure of the set $\Omega_{i}$),
$\Omega_{1}\cap\Omega_{2}=\emptyset$ and $\operatorname{supp}(f)\subset\Omega_{1}$,
i.e. all the space charge is contained in $\Omega_{1}$. The inner
domain $\Omega_{1}$, extending up to a given radius $R$, is our
computational domain and therefore must be selected to be as narrow
as possible.

Under this domain decomposition the problem~(\ref{eq:Initial problem})
turns into two coupled problems: 
\begin{equation}
\begin{array}{ccc}
\Delta\phi_{i}=f & \text{in} & \Omega_{i},\\
\phi_{i}=0 & \text{on} & \partial\Omega_{i}\setminus\Gamma,\\
\phi_{i}=\phi_{\Gamma} & \text{on} & \partial\Gamma,
\end{array}\label{dd}
\end{equation}
where $i=1,2$ and $\Gamma=\partial\Omega_{1}\cap\partial\Omega_{2}$
is the cylindrical surface at $\rho=R$ that separates the two domains.

Since at the interface $\Gamma$ both, $\phi_{1}$ and $\phi_{2}$,
are equal to the boundary value $\phi_{\Gamma}$, they fulfill $\phi_{1}=\phi_{2}$.
But besides this condition, in order for $\phi_{1}$ and $\phi_{2}$
to be consistent with the solution $\phi$ of the original problem~(\ref{eq:Initial problem}),
they must also satisfy 
\begin{equation}
\frac{\partial\phi_{1}}{\partial\rho}=\frac{\partial\phi_{2}}{\partial\rho}\;\text{on}\;\Gamma.\label{normals}
\end{equation}

\subsection{Linearity}
The linearity of the Poisson problems (\ref{dd}) with respect to
their sources $f$ allows us to decompose the potentials as 
\begin{equation}
\phi_{i}=\bar{\phi}_{i}[\phi_{\Gamma}]+\tilde{\phi}_{i}[f_{i}],
\end{equation}
where $\bar{\phi}_{i}[\phi_{\Gamma}]$ results from the boundary values
$\phi_{\Gamma}$ at the interface $\Gamma$ and $\tilde{\phi}_{i}[f_{i}]$
results from the original sources $f$ restricted
to $\Omega_{i}$ (we use $[\cdot]$ to denote
a functional dependence). The precise definitions read 
\begin{equation}
  \begin{array}{ccc}
    \Delta\bar{\phi}_{i}=0 & \text{in} & \Omega_{i},\\
    \bar{\phi}_{i}=0 & \text{on} & \partial\Omega_{i}\setminus\Gamma,\\
    \bar{\phi}_{i}=\phi_{\Gamma} & \text{on} & \partial\Gamma,
  \end{array}
  \label{barphi}
\end{equation}
and 
\begin{equation}
  \begin{array}{ccc}
    \Delta\tilde{\phi}_{i}=f & \text{in} & \Omega_{i},\\
    \tilde{\phi}_{i}=0 & \text{on} & \partial\Omega_{i}\setminus\Gamma,\\
    \tilde{\phi}_{i}=0 & \text{on} & \partial\Gamma.
  \end{array}
  \label{tildephi}
\end{equation}

In terms of these components the flux equation (\ref{normals}) can
be expressed as 
\begin{equation}
  \frac{\partial\bar{\phi}_{1}}{\partial\rho}
  \left[\phi_{\Gamma}\right] - \frac{\partial\bar{\phi}_{2}}{\partial\rho}
  \left[\phi_{\Gamma}\right] = 
  -\frac{\partial\tilde{\phi}_{1}}{\partial\rho}\left[f\right]\;
  \text{on}\;\Gamma,
\label{deconormals}
\end{equation}
where on the right hand side we have made use of $\tilde{\phi}_{2}=0$, since f=0 in $\Omega_2$.
\subsection{Expansion in orthonormal solutions of the Laplace equation}

The potentials $\bar{\phi}_{i}$ in (\ref{barphi}) are solutions
of the Laplace equation in cylindrical geometry and they can be expanded
using an orthogonal basis of solutions (see e.g. \cite{Jackson1975/book}):
\begin{subequations} 
\begin{equation}
  \bar{\phi}_{1} = \sum_{m=1}^{\infty}\alpha_{m}I_{0} \left(k_{m}\rho\right) \sin\left(k_{m}z\right),
\end{equation}
\begin{equation}
\bar{\phi}_{2}=\sum_{m=1}^{\infty}\beta_{m}K_{0}\left(k_{m}\rho\right)\sin\left(k_{m}z\right),
\end{equation}
\label{alphabeta} 
\end{subequations}
where $\alpha_{m}$ and $\beta_{m}$
are expansion coefficients, $k_{m}=m\pi/L$ and $I_{n}\left(x\right)$
and $K_{n}\left(x\right)$ are the modified $n$-order Bessel functions
of the first and second kind respectively. Note that the set $S=\left\{ \sin\left(k_{m}z\right)\right\} _{m=0}^{\infty}$
is an orthogonal basis of 
\begin{equation}
  \mathcal{L}^{2}\left(\left[0,L\right]\right)=
  \left\{ f:\left[0,L\right] \mapsto \mathbb{R}: 
    \int\left|f\left(z\right)\right|\text{\texttwosuperior}dz<\infty\right\},
  \label{eq:squareinteg}
\end{equation}
therefore, $\phi_{\Gamma}$ can be expanded as: 
\begin{equation}
  \phi_{\Gamma} \left(z\right) = \sum_{m=1}^{\infty }a_{m} \sin\left(k_{m}z\right).
\label{phigseries}
\end{equation}
If $\phi_{\Gamma}$ is continuous and piecewise differentiable
on $\left[0,L\right]$, 
$\phi_{\Gamma}'\in\mathcal{L}^{2}\left(\left[0,L\right]\right)$ and $\phi_{\Gamma}$ satisfies homogeneous Dirichlet boundary conditions,
then the sine series converges to $\phi_{\Gamma}$ uniformly on $\left[0,L\right]$.
Note that the term with $m=0$ vanishes due to the homogeneous boundary conditions
at $z=0$ and $z=L$.

The boundary conditions at $z=0$ and $z=L$ restrict the basis of solutions. Homogeneous Dirichlet boundary conditions
are simpler because there is only need for sine functions. However, if we had some other boundary conditions such as 
homogeneous Neumann, the convenient basis should also include cosine functions to allow for non-zero values of the potential
at $z=0$ and $z=L$. Nevertheless, this basis is not orthogonal and this would make things slightly more complicated.

\subsection{Continuity of the normal derivative}

Imposing that $\bar{\phi}_{1}=\bar{\phi_{2}}=\phi_{\Gamma}$ at $\rho=R$
we solve for $\alpha_{m}$ and $\beta_{m}$ and write (\ref{alphabeta})
as \begin{subequations} 
\begin{equation}
 \bar{\phi}_{1} = \sum_{m=1}^{\infty} a_{m} 
 \frac{I_{0}\left(k_{m}\rho\right)}{I_{0}\left(k_{m}R\right)}
 \sin\left(k_{m}z\right),
\label{barphi1}
\end{equation}
\begin{equation}
  \bar{\phi}_{2} = \sum_{m=1}^{\infty} a_{m} 
  \frac{K_{0}\left(k_{m}\rho\right)}{K_{0}\left(k_{m}R\right)}
  \sin\left(k_{m}z\right).
\end{equation}
\end{subequations} Using these expressions into the equation for
the normal derivatives (\ref{deconormals}) we obtain 
\begin{equation}
  \sum_{m=1}^{\infty} a_{m} k_{m} \left[
    \frac{I_{1}\left(k_{m}R\right)}{I_{0}\left(k_{m}R\right)} + 
    \frac{K_{1}\left(k_{m}R\right)}{K_{0}\left(k_{m}R\right)}
  \right]
  \sin(k_{m}z) = 
  -\left.\frac{\partial\tilde{\phi}_{1}}{\partial\rho}\right|_{\rho=R},
\end{equation}
where we have made use of the identities $I_{0}'(x)=I_{1}(x)$, $K_{0}'(x)=-K_{1}(x)$.
Using now the orthogonality of the basis $S$ we obtain equations
for the coefficients $a_{m}$: 
\begin{equation}
  \frac{L}{2}k_{m}a_{m}
  \left[
    \frac{I_{1}\left(k_{m}R\right)}{I_{0}\left(k_{m}R\right)} + 
    \frac{K_{1}\left(k_{m}R\right)}{K_{0}\left(k_{m}R\right)}
  \right]
  = -\int_{0}^{L}\dif z\, \sin\left(k_{m}z\right) \left. 
    \frac{\partial\tilde{\phi}_{1}}{\partial\rho}\right|_{\rho=R}.
\end{equation}

In a space discretization based on a cartesian grid
the integral in the latest expression is approximated
by a finite sum with the form of a Discrete Sine Transform (DST). This leads
to this final expression for the coefficients $a_{m}$ 
\begin{equation}
  a_{m} = -\frac{2}{m\pi}
  \left[
    \frac{I_{1}\left(k_{m}R\right)}{I_{0}\left(k_{m}R\right)} + 
    \frac{K_{1}\left(k_{m}R\right)}{K_{0}\left(k_{m}R\right)}
  \right]^{-1}\;
  \sum_{i=1}^{N}h \sin \left(k_{m}z_{i}\right) 
  \left.
    \frac{\partial\tilde{\phi}_{1}}{\partial\rho}
  \right|_{\rho=R,z=z_{i}} + \mathcal{O}(h^{2}),
  \label{am}
\end{equation}
where $h$ is the grid size and $\left\{ z_{i}\right\} _{i=1}^{N}$
are the solution nodes in the $z$-direction. In a discrete problem
the series in (\ref{phigseries}) is also truncated above $m=N$.

\subsection{Algorithm}
\label{sect:algorithm}
We are now ready to detail the domain-decomposition algorithm that
allows us to solve the Poisson equation in the reduced computational
domain $\Omega_{1}$ with free boundary conditions: 
\begin{enumerate}
\item Solve the Poisson equation in $\Omega_{1}$ with the source term $f$
and homogeneous Dirichlet boundary conditions at the boundary $\Gamma$.
Call the result $\tilde{\phi}_{1}$. 
\item Calculate the normal derivative of $\tilde{\phi}_{1}$ at $\Gamma$.
Apply a DST and use expression (\ref{am})
to obtain the coefficients $a_{m}$. 
\item Use these coefficients to obtain the boundary values $\phi_{\Gamma}$
by means of a second DST and expression (\ref{phigseries}). 
\item Solve again the Poisson equation in $\Omega_{1}$ but now use $\phi_{\Gamma}$
as inhomogeneous Dirichlet boundary condition at $\Gamma$. The result,
$\phi_{1}$ is the solution of the Poisson equation with free boundary
conditions. 
\end{enumerate}
To this algorithm we add the following remarks: 
\begin{enumerate}
\item After obtaining the coefficients $a_{m}$ one is tempted to use (\ref{barphi1})
together with $\phi_{1}=\bar{\phi}_{1}+\tilde{\phi}_{1}$ to avoid
solving the Poisson equation a second time. However, in a grid of
$M\times N$ cells this procedure takes about $MN^{2}$ operations whereas solving
the Poisson equation requires only $MN\log(MN)$ or $MN$ operations. 
\item The computational domain $\Omega_{1}$ has to be as narrow as possible
in order to reduce the computational cost of the simulation. Of course
this narrowing is limited by the constraint that $\Omega_{1}$ 
contains the support of the space charge density. In an electrostatic
discharge the charge density typically decays smoothly away from the
channel so in some cases one has to decide at which level it is safe
to truncate the charge density with an acceptable error. Nevertheless,
given the fast decay of the charge away from the channel, this is
probably not a serious concern in most cases. 
\end{enumerate}

\section{Tests and sample implementation}

\subsection{Tests}

In order to test our scheme we consider now a simple setup where 
the Poisson equation has a closed-form solution. 
An example of such a configuration is a uniformly
charged sphere located between two grounded, infinite planar electrodes.
The electrostatic potential in this setup can be calculated
by the method of images (see e.g. \citep{Jackson1975/book}) and equals
the potential created in free space by an infinite series of spheres
with alternating charges.

Suppose a sphere centred at $(\rho,z)=(0,z_{0})$ with radius $a<\min(z_{0},L-z_{0})$
and total charge $Q$. At a point with cylindrical coordinates $(\rho,z)$
the potential reads \begin{subequations} 
\begin{equation}
\phi(\rho,z)=\phi_{0}(\rho,z)+\frac{Q}{4\pi\epsilon_{0}}\sum_{\substack{k=-\infty\\
k\neq0
}
}^{\infty}\frac{(-1)^{k}}{\left[\rho^{2}+\left(z-z_{0}-2k(L-z_{0})\right)^{2}\right]^{1/2}},
\end{equation}
with 
\begin{equation}
\phi_{0}(\rho,z)=\frac{Q}{4\pi\epsilon_{0}}\begin{cases}
\frac{1}{\left[\rho^{2}+\left(z-z_{0}\right)^{2}\right]^{1/2}} & \text{if \ensuremath{\rho^{2}+(z-z_{0})^{2}>a^{2}}},\\
\frac{3a^{2}-\rho^{2}-(z-z_{0})^{2}}{2a^{3}} & \text{if \ensuremath{\rho^{2}+(z-z_{0})^{2}\leq a^{2}}}.
\end{cases}
\end{equation}
\label{images} \end{subequations}

\begin{figure}
\includegraphics[width=0.9\columnwidth]{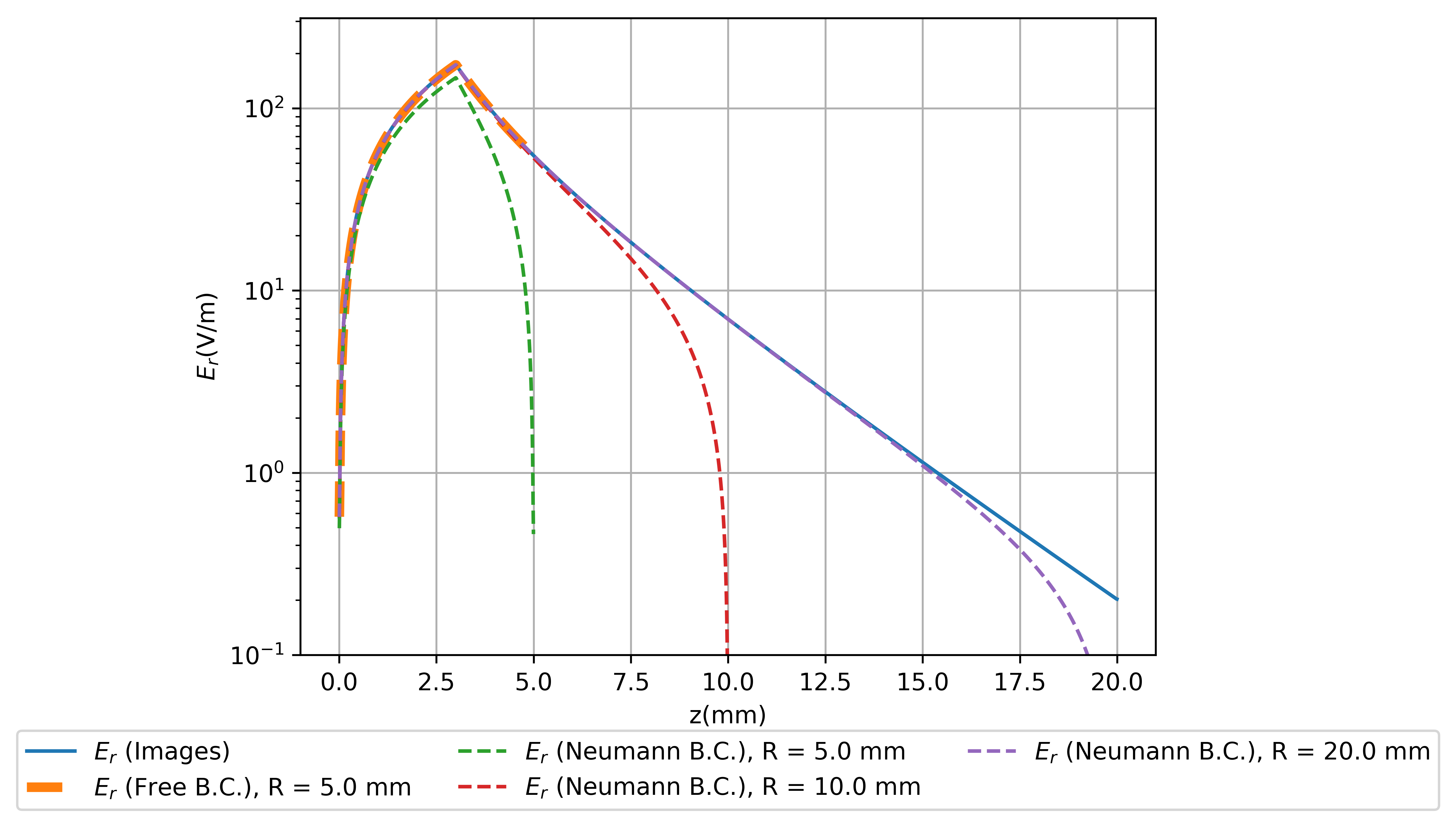} \caption{\label{fig:sphere} Comparison between electric fields created by
a uniformly charged sphere between two planar infinite electrodes
calculated by the approach presented in this work, by the method of
images and by imposing homogeneous Neumann boundary conditions at
the external boundary of the computational domain. Note the overlap
between the lines corresponding to free boundary conditions and to
the method of images.}
\end{figure}

Figure~\ref{fig:sphere} shows a comparison between the electric
fields computed using expression (\ref{images}) and using the approach
described in section~\ref{sect:description}. Here we took $a=\SI{3}{mm}$,
$L=\SI{10}{mm}$, $Q=\SI{e13}{e}$ (e is the elementary charge), 
$z_{0}=L/2$. For the
discretized solution we used $\Delta r=\Delta z=\SI{e-2}{mm}$
and a radial extension of the computational domain $R=\SI{5}{mm}$.
We also include the electrostatic potential calculated by imposing
homogeneous Neumann boundary conditions at the external boundary.

We see that the field calculated with the approach presented here
is indistinguishable from the field from the method of images. The
homogeneous Neumann conditions, on the other hand, produce an electric
field that at the surface of the sphere deviates by about 15\% from
the other two in the worst case, i.e. with $R=\SI{5}{mm}$. To investigate the convergence of the homogeneous Neumann
solution we extended the computational domain by computing the field
also for $R=\SI{10}{mm}$ and $R=\SI{20}{mm}$. As we move the external boundary away, the solution with Neumann conditions approaches our reference solution (Method of Images). Essentially, bringing the external boundary closer to the charged sphere shields the electric field before so the Neumann condition is fulfilled. 
As we will see, applied to streamer simulations, this leads to slightly lower values of the electric field in the streamer head and therefore less ionization. 

\subsection{Order of accuracy}
We have checked that the method described above does not change the order of accuracy
of the discretization of the Poisson equation by constructing a
closed-form solution of the Poisson equation that satisfies homogeneous 
Dirichlet boundary conditions in the upper and lower electrodes. We have used
the potential
\begin{equation}
\phi = \sin\left(\pi\frac{z}{L}\right)e^{-\frac{r^{2}}{\sigma^{2}}-\frac{\left(z-z_{0}\right)^{2}}{\sigma^{2}}},
\label{exact}
\end{equation}
whose Laplacian has the form
\begin{equation}
\begin{split}
\Delta \phi = &  \frac{1}{L^{2}\sigma^{4}}\left\lbrace 4 L \pi \sigma^{2} \left(z-z_{0}\right) \cos\left(\pi \frac{z}{L}\right) 
+ \left[\pi^{2}\sigma^{4}+L^{2}\left(-4r^{2}+6\sigma^{2}-4\left(z-z_{0}\right)^{2}\right)\right]\right\rbrace \\
& \times \sin\left(\pi\frac{z}{L}\right) 
e^{-\frac{r^{2}}{\sigma^{2}} - \frac{\left(z-z_{0}\right)^{2}}{\sigma^{2}}}.
\end{split}
\label{error}
\end{equation}

Although this charge density is not strictly bounded,
the contribution of charges excluded from the domain decays super-exponentially
as the domain becomes wider and can thus be neglected as long as the 
external radius of the computational domain is significantly longer than 
$\sigma$.

We have solved the Poisson equation corresponding to the Laplacian (\ref{error}) with
$L=\SI{1}{m}$, $\sigma=\SI{0.1}{m}$ and $z_0=\SI{0.5}{m}$ within a cylindrical domain 
with a radius $R=\SI{0.5}{m}$, where we imposed free boundary conditions with the 
method described above.  In this manner we checked that the convergence in the $\ell^2$-norm is of second order, the same as that of the finite 
difference scheme.  This is as expected because $\tilde{\phi}_{1}$, its derivative in 
the radial direction and the Fourier coefficients (\ref{am}) 
retain convergence of order $\mathcal{O}\left(h^{2}\right)$.

We are also interested in the convergence as we move the outer boundary. 
Following the example of the previous section, this time we change
the radius of the sphere to 0.1 mm and
the mesh spacing to $\SI{1} {\mu m}$. Errors are presented in Table \ref{tab:errors}, 
and the convergence is as expected of second order. Therefore, the decomposition
method does not cause errors of order less than two.

\begin{table}[H]
\centering{}%
\begin{tabular}{ccc}
Outer radius (mm) & $\left\Vert \epsilon\right\Vert _{2}$ & $\frac{\left\Vert \epsilon\right\Vert _{2}}{\left\Vert \phi_{exact}\right\Vert _{2}}$ \tabularnewline
\hline 
0.2 & \num{5.919e-7} & \num{5.037e-6}\tabularnewline
0.5 & \num{4.735e-8} & \num{3.812e-7}\tabularnewline
1 & \num{1.227e-8} & \num{9.838e-8}\tabularnewline
\end{tabular}\caption{Error obtained with change in outer radius}
\label{tab:errors}
\end{table}

\subsection{Sample implementation}

A computer code that produces a figure similar to 
figure~\ref{fig:sphere} is included
with this work. The code is implemented in Python and to be executed
it requires only the widely available scientific libraries NumPy and
SciPy. To solve the discrete Poisson equation the code constructs
a sparse matrix for the discrete Laplacian operator and invokes UMFPACK
\cite{Davis2004} (via SciPy) to solve the resulting linear system.

The code consists of a single file \verb|poisson_sparse_fft.py| and
contains the following methods: 
\begin{description}
\item{\verb|compute_matrix|:} Calculates the sparse matrix for
the discrete Laplacian operator in a given cartesian grid
and boundary conditions. 
\item{\verb|apply_inhom_bc|:} Modifies the right-hand-side of the
linear system in order to apply inhomogeneous Dirichlet boundary conditions. 
\item{\verb|DDM|:} Applies the domain decomposition method described
in section~\ref{sect:description} to solve the Poisson equation
with free boundary conditions. 
\item{\verb|MOI|:} Calculates the electrostatic potential by means
of the method of images, using (\ref{images}). 
\item{\texttt{main}:} This is the entry-point of the code: it uses
the above methods to produce the output figure. 
\end{description}

\section{Streamer simulations}
\label{sect:streamers}
\begin{figure}
\includegraphics[width=0.9\columnwidth]{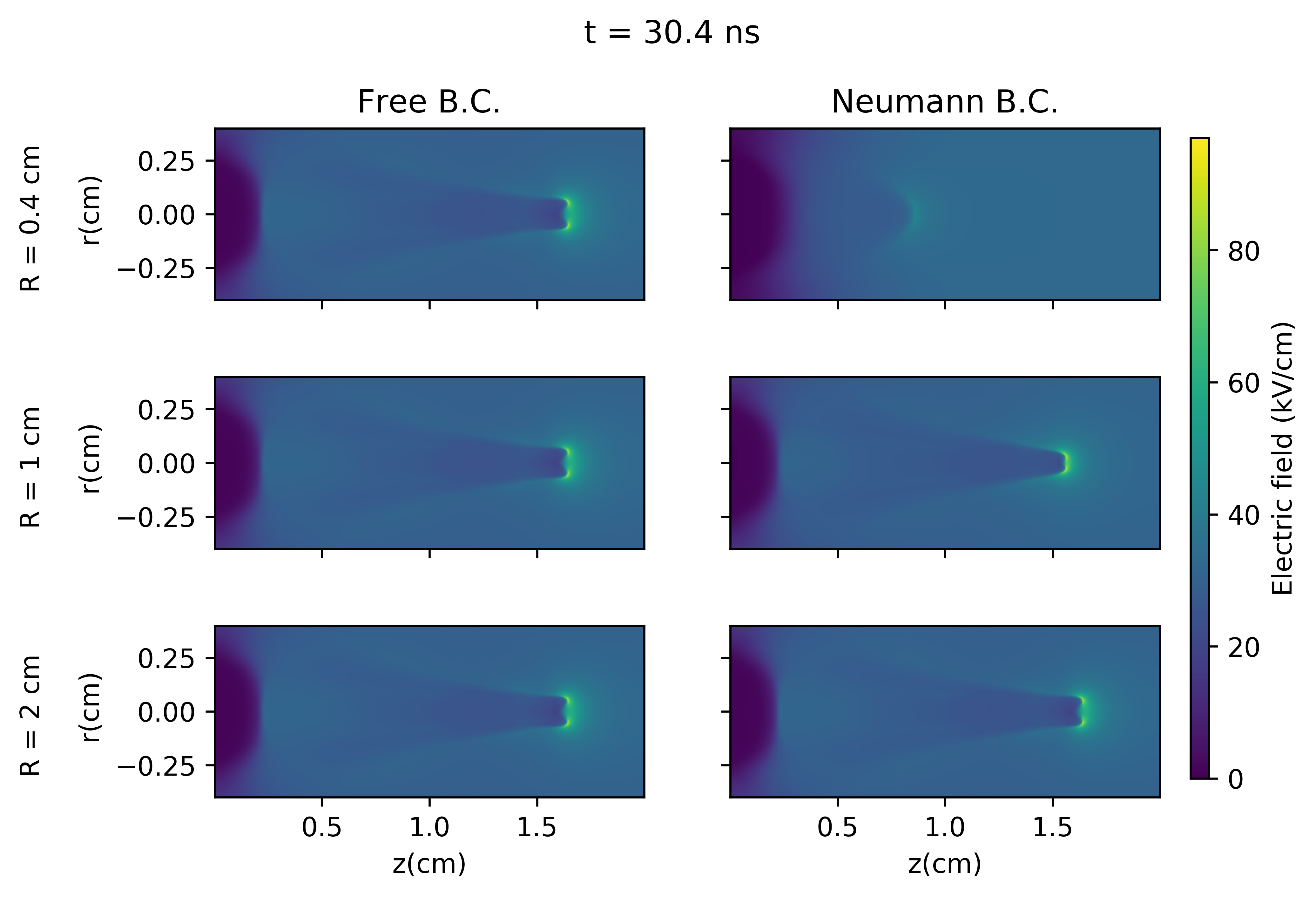} 
\caption{\label{fig:streamers} 
  Streamer simulations using free boundary conditions
  (left column) and homogeneous Neumann boundary conditions (right column)
  in the external boundary of the computational domain. For each selection
  of the boundary conditions we show three simulations where the computational
  domain extends to a radius $R=\SI{0.5}{cm}$ (top), $R=\SI{1}{cm}$
  (middle) and $R=\SI{2}{cm}$ (bottom).}
\end{figure}

In elongated electric discharges, Neumann boundary conditions are often considered more appropriate than Dirichlet to be applied at $r_{max}$  because there is not a physical electrode in the radial direction, and therefore there is no reason to keep constant the potential there.
The development of electric discharges is driven by long range interactions and therefore boundary conditions certainly affect the solution inside the computational domain. These effects can be reduced by enlarging the domain in the radial direction in exchange of a higher computational cost.
The procedure we have described allows us to keep the boundary $r_{max}$ close to the core of the simulation without noticeable numerical effects on the electric discharge. The following simulations clearly illustrate the features mentioned.

We simulated the propagation of streamer discharges between two planar
electrodes with a model that includes electron drift, impact ionization
and dissociative attachment 
and is implemented using the CLAWPACK/PetClaw library 
\cite{LeVeque2002/book,Alghamdi2011/PSP}.  The Poisson equation is solved
using the Improved Stabilized version of BiConjugate Gradient solver from the PETSc numerical library \cite{PETSc2016/web,PETSc2016/user}. For more
details about the physical model see e.g. references 
\citep{Montijn2006/JCoPh,Ebert2010/JGRA,Luque2012/JCoPh}.

We selected an inter-electrode gap of $L=\SI{2}{cm}$ and a background electric field of $\SI{27}{kV/cm}$. The
streamer is initiated by a neutral ionization seed attached to the
electrode on the central axis and centred at $z=L$. The peak electron
density in this seed is \SI{e14}{cm^{-3}}
and the $e$-folding length is \SI{0.7}{mm}.

As we are interested in the effect of the external boundary conditions,
we run simulations both with free boundary conditions, implemented
as described above, and with homogeneous Neumann conditions for the
electrostatic potential (as mentioned above,
it is generally assumed that Neumann boundary 
conditions introduce slightly smaller artifacts).
We also use different radii of the computational
domain, $R=\SI{0.5}{cm}$, $R=\SI{1}{cm}$ and $R=\SI{2}{cm}$. In
figure~\ref{fig:streamers} we show snapshots of the electric field
resulting from these simulations at time $t=\SI{30}{ns}$, shortly
after the streamer branches in the simulations with free boundary
conditions. Note however that the cylindrical symmetry of the simulations
prevents proper branching. MovieS1 (available online) shows the complete evolution of the streamers.

In the plots we see that the simulations with free boundary conditions
are essentially identical regardless of the lateral extension of the
computational domain. The simulations with homogeneous Neumann conditions
on the other hand depend artificially on the radius of the computational
domain. The streamer barely develops with $R=\SI{0.5}{cm}$ and only
the simulation with $R=\SI{2}{cm}$ reproduces accurately the branching
time of the simulations with free boundary conditions. We conclude
that, even in this case where the aspect ratio of the discharge is
not extremely high, the computational gain from reducing the domain
size (roughly a factor 4) more than compensates for the cost of solving
twice the Poisson equation, resulting in an overall improvement of
about a factor 2.

\section{Discussion and conclusions}
When they are not laterally constrained, most electrical discharges develop as 
elongated channels.  Despite different physical conditions and ionization 
mechanisms this feature is common to streamers, leaders and arcs.  The
underlying reason for this shared property is that all these processes
are affected by a Laplacian instability \cite{Arrayas2002/PhRvL,Derks2008/JNS},
whereby small
bumps in a discharge front enhance the electric field ahead 
and thus grow faster than the surrounding regions.  This prevents the 
formation of wide, smooth discharges and creates branched 
discharge trees of many filaments \cite{Luque2014/NJPh}.  

Since this is the preferred shape of a discharge, it is reasonable to optimize
our numerical models for elongated channels, selecting high-aspect-ratio 
computational domains.  The method and the code that we have presented here
can be used to achieve this efficiently and without losing accuracy.

We mention several possible extensions and refinements of this method.  
The first one, which is
described in the appendix, consists in extending the free boundary conditions 
also to the upper and lower simulation boundaries.  This can be useful for
the investigation of discharges not attached to any electrode or, with appropriate modifications, attached to a single electrode.  

A second extension is to adapt the method to run in parallel in several processors.  If we parallelize the Poisson solver by vertically decomposing the domain, the application of the method described above requires collecting information about the initial solution around the external boundary and then performing a one-dimensional Fourier transform.  The overhead of these steps is small compared to the operations required for the solution of the Poisson equation so the method can be efficiently parallelized.

Finally, one may ask about the suitability of this method for non-uniform meshes
and, in particular, for adaptively refined meshes.  Although the application of the method is in principle straightforward, a careful analysis
is required to understand the error incurred due to a possibly coarser resolution around the boundary than around a localized charge density.  This analysis, however, falls out of the scope of the present paper.

Note that although we have focused on the solution of the Poisson equation,
this method can be easily generalized to other elliptic partial differential
equations.  This can then applied to other components of streamer 
simulation codes such
as the speeding-up of photoionization calculations by approximating the 
interaction integral by combining solutions of a set of partial 
differential equations, as proposed
in references \cite{Segur2006/PSST,Luque2007/ApPhL,Bourdon2007/PSST}.

\appendix
\section{Full free boundary conditions}
\label{sect:fullfree}
The focus in this paper is the implementation of free boundary 
conditions in the outer boundary of a discharge confined between two
parallel plates but the method described above can also be extended to implement
free boundary conditions in all boundaries of a simulation with 
cylindrical symmetry.  In this appendix we describe this extension.

\subsection{Domain decomposition}
\begin{figure}[H]
\includegraphics[width=0.75\columnwidth]{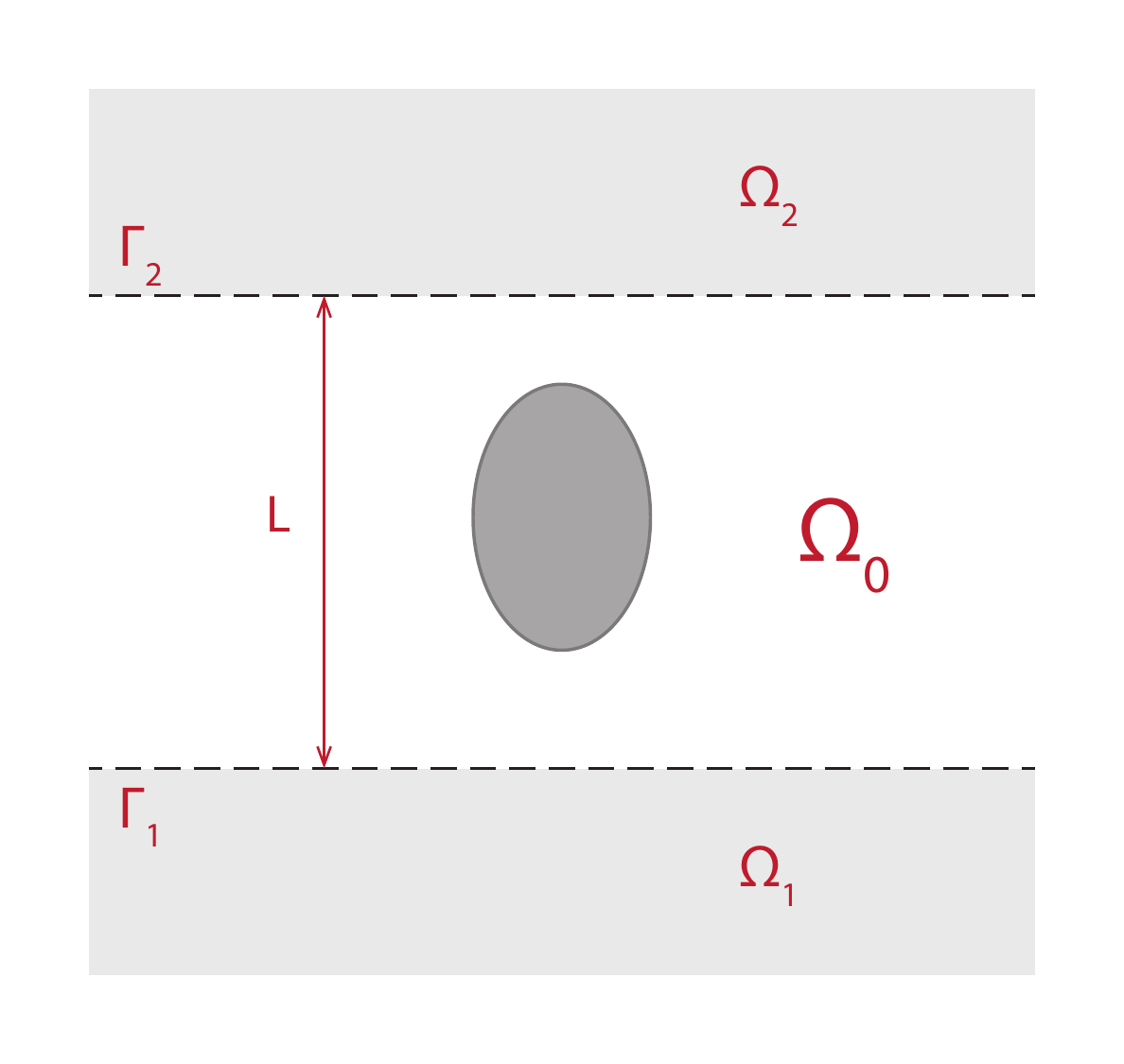} 
\caption{\label{fig:ddmschemefull} 
 Geometry of the problem. $\Omega$ is divided into three
subdomains: $\Omega_{0}$, $\Omega_{1}$ and $\Omega_{2}$.
The three extend indefinitely outwards and the latter two also downwards and upwards respectively.  }
\end{figure}

The method described above allows us to solve the Poisson equation in the space
between two infinite, parallel planes.  To build upon this procedure we decompose
now the full-space domain  $\Omega$ into three disjoint subdomains,
which we name $\Omega_{0}$, $\Omega_{1}$ and $\Omega_{2}$ (see 
figure~\ref{fig:ddmschemefull}).  We assume now that the support of the charge
distribution $f$ is contained in $\Omega_{0}$ and thus arrive at the three 
coupled problems
\begin{subequations}
\begin{equation}
\begin{array}{ccc}
\Delta\phi_{0}=f & \text{in} & \Omega_{0},\\
\phi_{0}=\phi_{\Gamma_{0i}} & \text{on} & \partial\Gamma_{0i}, i = 1, 2,
\end{array}
\end{equation}
and
\begin{equation}
\begin{array}{ccc}
\Delta\phi_{i}=f & \text{in} & \Omega_{i},\\
\phi_{i}=\phi_{\Gamma_{0i}} & \text{on} & \partial\Gamma_{0i},
\end{array}
\end{equation}
\label{ddfull}
\end{subequations}
where $i=1,2$ and $\Gamma_{0i}=\partial\Omega_{0}\cap\partial\Omega_{i}$
are the surfaces $z=0$ and $z=L$ respectively.

Since there are two interfaces, there are also two conditions for the continuity
of the normal derivative:
\begin{equation}
\frac{\partial\phi_{0}}{\partial z}=\frac{\partial\phi_{i}}{\partial z}\;\text{on}\;\Gamma_i,i=1,2.\label{normalsfull}
\end{equation}

\subsection{Linearity}
Linearity allows us to split the problem into 
\begin{subequations}
\begin{equation}
  \begin{array}{ccc}
    \Delta\bar{\phi}_{0}=0 & \text{in} & \Omega_{0},\\
    \bar{\phi}_{0}=\phi_{\Gamma_{0i}} & \text{on} & \partial\Gamma_{0i}, i = 1,2,
  \end{array}
  \label{barphifull0}
\end{equation}
\begin{equation}
  \begin{array}{ccc}
    \Delta\tilde{\phi}_{0}=f & \text{in} & \Omega_{0},\\
    \tilde{\phi}_{0}=0 & \text{on} & \partial\Gamma_{0i}, i = 1, 2
  \end{array}
  \label{tildephifull0}
\end{equation}
\end{subequations}
and for $i = 1, 2$,
\begin{subequations}
\begin{equation}
  \begin{array}{ccc}
    \Delta\bar{\phi}_{i}=0 & \text{in} & \Omega_{i},\\
    \bar{\phi}_{i}=\phi_{\Gamma_{0i}} & \text{on} & \partial\Gamma_{0i},
  \end{array}
  \label{barphifull}
\end{equation}
\begin{equation}
  \begin{array}{ccc}
    \Delta\tilde{\phi}_{i}=f & \text{in} & \Omega_{i},\\
    \tilde{\phi}_{i}=0 & \text{on} & \partial\Gamma_{0i}.
  \end{array}
  \label{tildephifull}
\end{equation}
\end{subequations}

Since there is no charge outside the computational domain, $\tilde{\phi}_i = 0$
for $i=1,2$.
These equations are naturally subject to the condition that the potential 
vanishes at $-z,z,\rho\rightarrow\infty$.

Note now that the problem (\ref{tildephifull0}) can be solved by 
the procedure described in the main text, since $\tilde{\phi}_{0}$ is the 
union of the solutions to the problems at (\ref{dd}).

In terms of these components, the flux equation (\ref{normals}) can be expressed as 
\begin{equation}
  \frac{\partial\bar{\phi}_{0}}{\partial z}
  \left[\phi_{\Gamma_{0i}}\right] - \frac{\partial\bar{\phi}_{i}}{\partial z}
  \left[\phi_{\Gamma_{0i}}\right] = 
  -\frac{\partial\tilde{\phi}_{0}}{\partial z}\left[f\right]\;
  \text{on}\;\Gamma,\; \forall i.
\label{deconormalsfull}
\end{equation}

\subsection{Expansion in solutions of the Laplace equation}

The potentials $\bar{\phi}_{i}$ in (\ref{barphifull}) are solutions
of the Laplace equation in cylindrical coordinates and they can be expanded 
using a set of solutions  (see e.g. \cite{Jackson1975/book}). Since the domain is unbounded in the radial direction, instead of a series expansion 
we obtain an integral transform, which we can write in terms of the zero-order 
Hankel transform, which reads:

\begin{subequations} 
\begin{equation}
  \bar{\phi}_{0} = \int_{0}^{\infty}k \dif k\left[A\left(k\right)e^{kz}+B\left(k\right)e^{-kz}\right]J_{0}\left(k\rho\right),
\end{equation}
\begin{equation}
	\bar{\phi}_{1}=\int_{0}^{\infty}k \dif k \:C\left(k\right)e^{kz}J_{0}\left(k\rho\right),
\end{equation}
\begin{equation}
	\bar{\phi}_{2}=\int_{0}^{\infty}k\dif k \:D\left(k\right)e^{-kz}J_{0}\left(k\rho\right),
\end{equation}
\label{abcd} 
\end{subequations} 
where $J_{0}\left(x\right)$ is the zero-order Bessel function of the first kind and the functions $A, B, C, D$ weight the independent solutions to 
the Laplace equation.  Note that, although the factor $k$ can be absorbed into these functions, it appears explicitly in order to show the Hankel transform 
structure.

The function $\phi_{\Gamma_{0i}}$ can also be transformed as:
\begin{equation}
  \phi_{\Gamma_{0i}} \left(\rho\right) = \int_{0}^{\infty}k \dif k E_{i}\left(k\right) J_{0}\left(k\rho\right).
\label{gammas}
\end{equation}

Casting these equations in the form of a Hankel transform is important because
as the Hankel transform can be inverted (it is its own inverse) we can use the 
fact that, subject to some regularity assumptions,

\begin{equation}
\int_{0}^{\infty}k\dif k \:F\left(k\right)J_{0}\left(k\rho\right) = 0 \iff F(k)=0.
\label{hankel}
\end{equation}

\subsection{Continuity of the normal derivative}

Imposing that $\bar{\phi}_{0}=\bar{\phi_{i}}=\phi_{\Gamma_{0i}}$ at $z=0\;\text{and}\;L,$
we solve for $A, B, C, D$ using (\ref{hankel}) and write (\ref{abcd})
as \begin{subequations} 
\begin{equation}
 \bar{\phi}_{0} = \int_{0}^{\infty}
 \frac{k\dif k}{e^{2kL} - 1}
 \left[ 
 \left(E_{2}e^{kL}-E_{1}\right)
 e^{kz} + \left(-E_{2}+E_{1}e^{kL}\right)
 e^{-k\left(z-L\right)}
 \right]
 J_{0}\left(k\rho\right),
\label{barphifull0exp}
\end{equation}
\begin{equation}
 \bar{\phi}_{1} = \int_{0}^{\infty}
 k\dif k\:E_{1}e^{kz}
 J_{0}\left(k\rho\right),
\label{barphifull1}
\end{equation}
\begin{equation}
 \bar{\phi}_{2} = \int_{0}^{\infty}
 k\dif k\:E_{2}e^{-k\left(z-L\right)}
 J_{0}\left(k\rho\right).
\label{barphifull2}
\end{equation}
\end{subequations}

Using these expressions into the equation for the
normal derivative (\ref{deconormalsfull}) we obtain
\begin{subequations}
\begin{equation}
 \int_{0}^{\infty}
 \frac{2e^{kL}k^{2}\dif k}{e^{2kL} - 1} 
 \left(E_{2}-E_{1}e^{kL}\right)
 J_{0}\left(k\rho\right) = -\left.\frac{\partial\tilde{\phi}_{0}}{\partial z}\right|_{z=0},
\end{equation}
\begin{equation}
 \int_{0}^{\infty}
 \frac{e^{kL}k^{2}\dif k}{e^{2kL} - 1} 
 \left(E_{2}e^{kL}-E_{1}\right)
 J_{0}\left(k\rho\right) = -\left.\frac{\partial\tilde{\phi}_{0}}{\partial z}\right|_{z=L}.
\end{equation}
\end{subequations}

We can obtain the coefficients $E_{1}$ and $E_{2}$ going back to the $k$-space by means of the Hankel transform:
\begin{subequations}
\begin{equation}
 E_{1}\left(k\right) = \frac{1}{2}\left(e^{-kL}I_{L} - I_{0}\right),
\end{equation}
\begin{equation}
 E_{2}\left(k\right) = \frac{1}{2}\left(I_{L} - e^{-kL}I_{0}\right),
\end{equation}
\label{E12}
\end{subequations}

where

\begin{equation}
I_{0,L}\left(k\right) = -\frac{1}{k} \int_{0}^{\infty}\rho \dif \rho \left.\frac{\partial\tilde{\phi}_{0}}{\partial z}\right|_{z=0,L}J_{0}\left(k\rho\right).
\label{I0IL}
\end{equation}

In this expression $\tilde{\phi}_{0}$ and its normal derivative are known 
from the algorithm described in \ref{sect:algorithm}.  Therefore we can compute
$I_{0,L}$ and, using (\ref{E12}) $E_{1,2}$.  These functions in turn can be 
inserted in (\ref{gammas}) to yield the boundary condition to impose on
$\Gamma_{1,2}$.  

\section*{Supplementary Data}

The Supplementary Data include a movie (MovieS1) showing the development of
several streamer discharges within domains of different radial extent
as described in section \ref{sect:streamers}.
Among a total of six cases, three of them show the
development of the discharge for Poisson's equation subject to Neumann boundary
conditions whereas the remaining three correspond to 
the solution of Poisson's equation with free radial boundary conditions. Figure \ref{fig:streamers} is a frame extracted from this 
movie.

\section*{Acknowledgement}

This work was supported by the European Research Council (ERC) under
the European Union H2020 programme/ERC grant agreement 681257.

\newcommand{\mdash}{---}
\newcommand{\nat}{Nature}
\newcommand{\pre}{Phys. Rev. E}
\newcommand{\prb}{Phys. Rev. B}
\newcommand{\physrep}{Phys. Rep.}
\bibliographystyle{elsarticle-num}
\bibliography{library}

\end{document}